\documentclass[conference,compsoc]{IEEEtran}
\ifCLASSOPTIONcompsoc
  \usepackage[nocompress]{cite}
\else
  \usepackage{cite}
\fi
\ifCLASSINFOpdf
\else
\fi
\usepackage{amsmath}
\usepackage{amssymb}
\usepackage{amsthm}
\usepackage{mathtools}
\usepackage{bm}
\usepackage{mathrsfs}
\usepackage{cases}
\usepackage{esvect}
\usepackage{siunitx}
\usepackage{algorithm}

\usepackage{graphicx}

\begin{document}
%
\title{Unveiling Patterns in European Airbnb Prices: A Comprehensive Analytical Study Using Machine Learning Techniques}

\author{\IEEEauthorblockN{Trinath Sai Subhash Reddy Pittala\IEEEauthorrefmark{1}}
\IEEEauthorblockA{School of Computing\\
Clemson University\\
Clemson, South Carolina 29631\\
Email: tpittal@g.clemson.edu}
\and
\IEEEauthorblockN{Uma Maheswara R Meleti\IEEEauthorrefmark{2}}
\IEEEauthorblockA{School of Computing\\
Clemson University\\
Clemson, South Carolina 29631\\
Email: umeleti@g.clemson.edu}
\and
\IEEEauthorblockN{Hemanth Vasireddy\IEEEauthorrefmark{2}}
\IEEEauthorblockA{School of Computing\\
Clemson University\\
Clemson, South Carolina 29631\\
Email: hvasire@g.clemson.edu}}


%


\maketitle

\begin{abstract}
In the burgeoning market of short-term rentals, understanding pricing dynamics is crucial for a range of stakeholders. This study delves into the factors influencing Airbnb pricing in major European cities, employing a comprehensive dataset sourced from Kaggle. We utilize advanced regression techniques, including linear, polynomial, and random forest models, to analyze a diverse array of determinants, such as location characteristics, property types, and host-related factors. Our findings reveal nuanced insights into the variables most significantly impacting pricing, highlighting the varying roles of geographical, structural, and host-specific attributes. This research not only sheds light on the complex pricing landscape of Airbnb accommodations in Europe but also offers valuable implications for hosts seeking to optimize pricing strategies and for travelers aiming to understand pricing trends. Furthermore, the study contributes to the broader discourse on pricing mechanisms in the shared economy, suggesting avenues for future research in this rapidly evolving sector.
\end{abstract}


%
\IEEEpeerreviewmaketitle

\section{Introduction}

In recent years, the shared economy has witnessed significant growth, with platforms like Airbnb transforming the landscape of short-term accommodations. Particularly in Europe, Airbnb has emerged as a popular alternative to traditional lodging options, catering to a diverse array of travelers seeking unique and personalized experiences. However, this rapid expansion has brought forth complex pricing dynamics, influenced by an array of factors ranging from location specifics to property characteristics. Understanding these pricing mechanisms is not only critical for hosts and guests but also provides valuable insights into market trends and consumer preferences within the shared economy.

This study aims to dissect the determinants of Airbnb pricing across major European cities, employing a data-driven approach. By leveraging a rich dataset from Kaggle (Gyódi \& Nawaro, 2021)\cite{gyodi2021determinants}, encompassing a variety of variables including location, room type, and host information, we apply sophisticated machine learning techniques, namely regression analysis and random forest models, to unearth underlying patterns and influences in Airbnb pricing. Our exploration extends beyond simple cost determinants, delving into how different factors interplay to shape pricing strategies in diverse urban contexts.

The significance of this research lies in its potential to guide Airbnb hosts in optimizing their pricing models, aiding travelers in making informed decisions, and contributing to the broader understanding of economic models in the shared economy. Additionally, this study serves as a testament to the power of data analytics in deciphering complex market phenomena, offering a blueprint for similar analyses in other domains.

\section{Literature Review}

The advent of Airbnb and similar platforms has catalyzed a paradigm shift in the hospitality sector, prompting a growing body of research focused on pricing strategies within the shared economy. Several studies have examined the determinants of pricing in short-term rental markets, highlighting the interplay of various factors such as location, property characteristics, and host reputation.

  \subsection{Location-Based Pricing Factors:} A significant amount of research has underscored the importance of location in Airbnb pricing. Studies by \cite{gibbs2018pricing} and \cite{chen2017consumer} have demonstrated that proximity to city centers and tourist attractions often leads to higher prices due to increased demand. Furthermore, regional variations, as explored by \cite{zhang2017key}, indicate that cultural and economic factors in different cities and countries can significantly influence pricing.

  \subsection{Property Characteristics and Amenities:} The impact of property features on pricing has been another area of focus. Research by \cite{perez2018and} and \cite{teubner2017price} has shown that factors such as the size of the property, the number of bedrooms and bathrooms, and the availability of luxury amenities like swimming pools are directly correlated with higher rental prices.

  \subsection{Host-Related Factors:} The influence of host attributes on Airbnb pricing has been explored in studies like those by \cite{abrate2022dynamic} and \cite{casamatta2022host}. These include factors such as the host's reputation, response rate, and the number of listings they manage, all of which have been found to affect pricing decisions and consumer choices.

  \subsection{Machine Learning in Price Prediction:} With the advent of big data and advanced analytics, machine learning models have increasingly been employed to predict Airbnb prices. The works of \cite{huang2022pricing} and \cite{hong2020analyzing} illustrate the efficacy of regression models and random forest algorithms in predicting rental prices, based on a multitude of variables.
  Despite this extensive research, there remains a gap in comprehensive, multi-city, European-focused studies that utilize advanced machine learning techniques to analyze Airbnb pricing. This study aims to bridge this gap by applying sophisticated data analytics tools to a diverse dataset, providing a nuanced understanding of pricing dynamics across major European cities.

\section{Methods}

This study employs a multifaceted analytical approach to investigate the determinants of Airbnb pricing in major European cities \textbf{(Amsterdam, Athens, Barcelona, Berlin, Budapest, Lisbon, London, Paris, Rome, Vienna)}. The methodology is underpinned by robust data analysis and machine learning techniques, specifically regression analysis and random forest models.

  \subsection{Dataset Description:} The analysis is based on the 'Airbnb Price Determinants' dataset from Kaggle, which includes data for 51,707 listings across ten major European cities. Key variables in the dataset include host information (e.g., host ID, response rate), property characteristics (e.g., room type, number of bedrooms, amenities), location details (e.g., latitude, longitude, proximity to city center), and pricing information.

    \subsection{Data Preprocessing:} The dataset underwent comprehensive preprocessing to ensure its integrity and applicability for analysis. The following key steps were meticulously executed:

    \subsubsection{Data Cleaning:} This crucial step involved the elimination of duplicate entries and the management of missing values to enhance data quality. A significant decision in this phase was the removal of \texttt{room\_shared} and \texttt{room\_private} variables due to their redundancy with the \texttt{room\_type} variable. This action was taken to address multicollinearity concerns, ensuring a more robust dataset for subsequent analysis.
    
    \subsubsection{Feature Engineering:} In this stage, we engineered new variables to deepen our analysis. This included the categorization of properties by size and type, offering a more nuanced view of the dataset. Additionally, we quantified properties' proximity to major attractions, acknowledging its potential impact on pricing.

    \subsubsection{Normalization:} To facilitate a fair comparison across various scales, we standardized numerical variables. This normalization process was vital for maintaining consistency and preventing skewed interpretations in the analysis.

    \subsubsection{Data Transformation:} The transformation phase involved converting categorical variables into dummy variables. This step was essential for their effective integration into our regression models, allowing for a comprehensive analysis that incorporates both quantitative and qualitative data.

  \subsection{Exploratory Data Analysis:} The exploratory data analysis of the dataset yielded insightful trends and patterns regarding Airbnb pricing across various European cities. The key findings from this analysis are as follows:

    \subsubsection{Pricing Trends Across Cities:} A notable trend emerged with Amsterdam exhibiting the highest average prices for Airbnb listings in Europe as shown in Figure 1. This suggests a unique market dynamic in Amsterdam that drives higher pricing compared to other cities.
    \begin{figure}[htbp]
      \centering
      \includegraphics[width=0.5\textwidth]{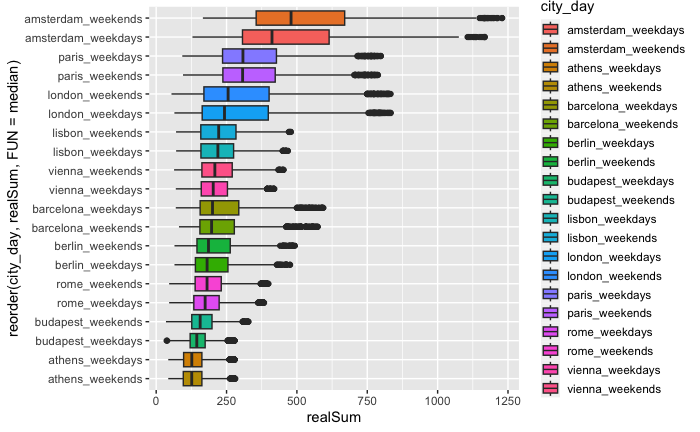}
      \caption{}
      \label{fig:my_label}
    \end{figure}
    \subsubsection{Impact of Property Type on Pricing:} Analysis indicates a clear distinction in pricing based on property type. Entire homes generally command higher prices compared to shared or private rooms, reflecting a premium placed on privacy and space by Airbnb users.
    \begin{figure}[htbp]
      \centering
      \includegraphics[width=0.5\textwidth]{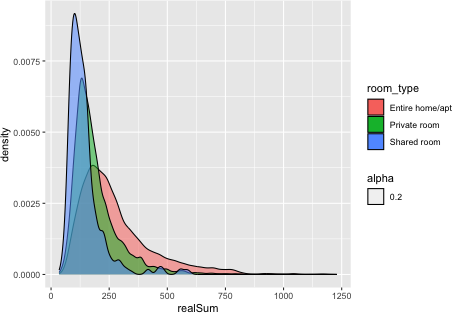}
      \caption{}
      \label{fig:my_label}
    \end{figure}
    \subsubsection{Proximity to Metro Stations and Pricing Correlation:} A common trend across most cities showed that properties located closer to metro stations tend to have higher prices, highlighting the value guests place on accessibility and convenience. However, an interesting anomaly was observed in Rome, where the distance to the nearest metro station appeared to have a uniform effect on pricing for all property types. This indicates that in certain urban contexts, other factors may override proximity to public transport as a primary pricing determinant.
    \begin{figure}[htbp]
      \centering
      \includegraphics[width=0.5\textwidth]{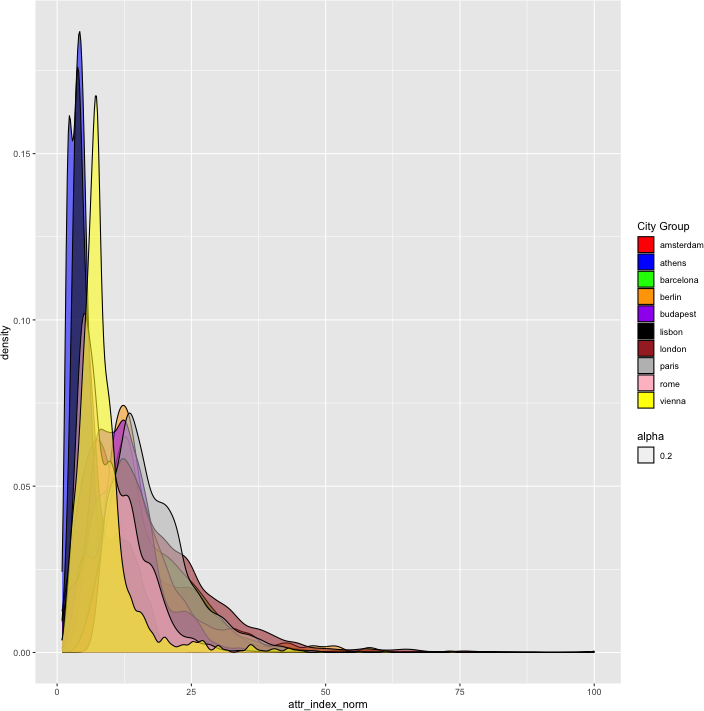}
      \caption{}
      \label{fig:my_label}
    \end{figure}
    \subsubsection{City Attractiveness Index and Its Influence:} The City Attractiveness Index significantly varies among European cities, with notable examples like Paris scoring higher. This variation directly correlates with differences in Airbnb room prices, highlighting the impact of a city’s overall appeal on accommodation pricing. Cities with higher attractiveness tend to have increased demand for accommodations, thus driving up prices. Factors contributing to this index include cultural landmarks, entertainment options, and general livability. For Airbnb hosts, this means that properties in more attractive cities can command higher prices. For travelers, it indicates a likely correlation between a city's appeal and accommodation costs. This index not only influences individual pricing strategies but also reflects broader economic and tourism trends within a city.
    \subsubsection{Price Disparity in Room Types:} The analysis highlighted a clear disparity in pricing between shared and non-shared rooms. Non-shared rooms, which include entire apartments or private rooms, generally command higher prices. This trend can be attributed to the additional privacy and exclusivity that these types of accommodations offer. This finding underscores the value guests place on having a private space, which becomes a significant factor in their accommodation choices and willingness to pay more. Understanding this disparity is crucial for Airbnb hosts in setting competitive prices and for guests in making informed accommodation decisions.

  \subsection{Regression Analysis:}

    \subsubsection{Multivariate Linear Regression:} Linear regression was implemented as the foundational model to establish a baseline understanding of the relationship between independent variables and Airbnb pricing. This approach assumes a linear relationship between the dependent variable (price) and the independent variables, such as location, property type, and host characteristics. The model's coefficients provide insights into the average effect of each predictor on the pricing, offering a preliminary overview of the factors that most significantly influence Airbnb prices. Here the realSum is predicted by all the other variables as shown in equation(1).
    \begin{equation}
      \begin{aligned}
      \text{M1} \leftarrow & \ \text{lm}(\text{realSum} \sim . - \text{realSum} - \text{X}, \\
      & \text{data} = \text{my\_data\_train})
      \end{aligned}
    \end{equation}
    \subsubsection{Polynomial Regression:} In addressing potential non-linear relationships between variables and pricing, polynomial regression was utilized. This approach expands upon linear regression by incorporating polynomial terms, enabling the model to flex and capture more intricate patterns in the data. By scrutinizing these non-linear effects, we attain a more profound comprehension of how variations in variables, particularly at various levels, influence Airbnb pricing. Equation (2) illustrates the prediction of realSum by all other nominal and ordinal variables in this context.
    \begin{equation}
      \begin{aligned}
      \text{poly3} \leftarrow & \ \text{lm}(\text{realSum} \sim \text{room\_type} \\
      & + \text{host\_is\_superhost} \\
      & + \text{multi} + \text{biz} + \text{city\_day} \\
      & + \text{person\_capacity} \\
      & + \text{cleanliness\_rating} \\
      & + \text{guest\_satisfaction\_overall} \\
      & + \text{bedrooms} \\
      & + \text{poly(dist, 3)} \\
      & + \text{poly(metro\_dist, 3)} \\
      & + \text{poly(attr\_index\_norm, 3)} \\
      & + \text{poly(rest\_index\_norm, 3)} \\
      & + \text{lng} + \text{lat}, \\
      & \text{data} = \text{my\_data\_train})
      \end{aligned}
    \end{equation}      
    \subsubsection{Interaction Variables:} The analysis encompassed an examination of interaction effects among variables to discern the compounded influences on pricing. Interaction terms were incorporated into the regression models to investigate how the association between an independent variable and price varies across different levels of another variable. This is especially vital for comprehending how combinations of factors, such as location and property type, collectively shape rental prices. In equation (3), realSum is predicted by all other variables, and interactions are computed specifically among continuous variables.
    \begin{equation}
      \begin{aligned}
      \text{M1IV} \leftarrow & \ (\text{realSum} \sim \text{room\_type} \\
      & + \text{host\_is\_superhost} \\
      & + \text{multi} + \text{biz} + \text{city\_day} \\
      & + \text{person\_capacity} \\
      & + \text{cleanliness\_rating} \\
      & + \text{guest\_satisfaction\_overall} \\
      & + \text{bedrooms} + \text{dist} \\
      & + \text{metro\_dist} \\
      & + \text{attr\_index\_norm} \\
      & + \text{rest\_index\_norm} \\
      & + \text{lng} + \text{lat} \\
      & + \text{metro\_dist} : \text{dist} \\
      & + \text{attr\_index\_norm} : \text{dist} \\
      & + \text{attr\_index\_norm} : \text{metro\_dist} \\
      & + \text{rest\_index\_norm} : \text{dist} \\
      & + \text{rest\_index\_norm} : \text{metro\_dist} \\
      & + \text{rest\_index\_norm} : \text{attr\_index\_norm}, \\
      & \text{data} = \text{my\_data\_train})
      \end{aligned}
    \end{equation}
    \subsubsection{Lasso Regression:} Lasso Regression was utilized as a method of regression that not only helps in predicting outcomes but also assists in feature selection. By imposing a penalty on the absolute size of the coefficients, Lasso tends to shrink less important predictor coefficients to zero, effectively performing variable selection. This is particularly beneficial in a dataset with many predictors, as it simplifies the model by keeping only the most relevant variables. Lasso's ability to perform regularization helps in avoiding overfitting, making it a robust tool for modeling and understanding the key drivers of Airbnb pricing.

  \subsection{Random Forest Model:} A random forest regression model was developed, building upon the insights gained from the linear regression analysis. 
  The choice of this model was particularly driven by its proficiency in managing the dataset's inherent complexity (by maintaining all non redundant variables as in equation (4)) and the non-linear relationships uncovered during the linear regression phase. Key aspects that informed this choice include:

    \subsubsection{Capability with Complex Data:} The random forest model is adept at handling datasets with a multitude of predictor variables. In our case, the diverse range of factors influencing Airbnb pricing - from location specifics to host-related attributes - necessitates a model capable of processing and making sense of this complexity.
    \begin{equation}
      \begin{aligned}
      \text{rf\_model} \leftarrow & \ \text{randomForest}(\text{realSum} \sim \text{room\_type} \\
      & + \text{host\_is\_superhost} \\
      & + \text{multi} + \text{biz} + \text{city\_day} \\
      & + \text{person\_capacity} \\
      & + \text{cleanliness\_rating} \\
      & + \text{guest\_satisfaction\_overall} \\
      & + \text{bedrooms} + \text{dist} \\
      & + \text{metro\_dist} \\
      & + \text{attr\_index\_norm} \\
      & + \text{rest\_index\_norm} \\
      & + \text{lng} + \text{lat} \\
      & \text{data} = \text{my\_data\_train})
      \end{aligned}
    \end{equation}
    \subsubsection{Non-Linearity Accommodation:} Linear regression analysis revealed the presence of non-linear patterns in the data, a scenario where random forest models excel. By aggregating the outputs of numerous decision trees, each considering a random subset of features, the model is able to capture and reflect these non-linear relationships effectively.
    \subsubsection{Robustness Against Overfitting:} A key strength of the random forest model is its inherent mechanism to prevent overfitting. The model achieves this by creating multiple decision trees and using their averaged predictions, which generally results in a more reliable and accurate model, particularly important in our analysis involving a large and complex dataset.
    \subsubsection{Handling Large Datasets:} Our study involves a comprehensive dataset encompassing a wide array of Airbnb listings across multiple European cities. The random forest model is well-suited for such large datasets, providing the computational efficiency and scalability needed to process and analyze extensive data effectively.

  \subsection{Model Evaluation Metrics:} Models were evaluated based on their R-squared values, mean squared error (MSE), and cross-validation scores to ensure accuracy and reliability.

  \begin{table*}[ht]
    \centering
    \caption{Summary of Regression and Random Forest Results}
    \label{tab:model_results}
    \begin{tabular}{|c|c|c|c|c|c|c|}
    \hline
    \textbf{Model} & \textbf{Train \(R^2\)} & \textbf{Train Adj \(R^2\)} & \textbf{Train \(RMSE\)} & \textbf{Test \(R^2\)} & \textbf{Test Adj \(R^2\)} & \textbf{Test \(RMSE\)} \\ \hline
    MLR & 0.2150414 & 0.2146725 & 304.9175 & 0.33744 & 0.3371287 & 233.2618 \\ \hline
    MLR with Step & 0.2149964 & 0.2146275 & 304.9262 & 0.3373441 & 0.3370327 & 233.2787 \\ \hline
    MLR with IVs & 0.2159815 & 0.215613 & 304.7348 & 0.3379444 & 0.3376333 & 233.173 \\ \hline
    MLR with IVs, Step & 0.2159435 & 0.215575 & 304.7422 & 0.3380392 & 0.3377281 & 233.1563 \\ \hline
    Poly with Order 2 & 0.2154699 & 0.2151012 & 304.8342 & 0.3373538 & 0.3370424 & 233.277 \\ \hline
    Poly with Order 2, Step & 0.2154289 & 0.2150602 & 304.8422 & 0.3373204 & 0.337009 & 233.2829 \\ \hline
    Poly with Order 2, IVs & 0.22214 & 0.2217744 & 303.5356 & 0.334993 & 0.3346805 & 233.6922 \\ \hline
    Poly with Order 2, IVs, Step & 0.2221344 & 0.2217689 & 303.5367 & 0.3350694 & 0.3347569 & 233.6788 \\ \hline
    Poly with Order 3 & 0.2160624 & 0.215694 & 304.7191 & 0.3375855 & 0.3372742 & 233.2362 \\ \hline
    Poly with Order 3, Step & 0.2160337 & 0.2156653 & 304.7247 & 0.3376519 & 0.3373407 & 233.2245 \\ \hline
    Poly with Order 3, IVs & 0.2330663 & 0.2327059 & 301.3962 & 0.1901115 & 0.189731 & 257.8955 \\ \hline
    Poly with Order 3, IVs, Step & 0.2285384 & 0.2281759 & 302.2846 & 0.3281442 & 0.3281442 & 234.8925 \\ \hline
    Lasso With Order 1 & 0.215693 & 0.2147166 & 325.7776 & 0.3372306 & 0.3353023 & 233.2987 \\ \hline
    Lasso With Order 2 & 0.2201963 & 0.217819 & 326.6683 & 0.1830879 & 0.1772535 & 259.0113 \\ \hline
    Lasso With Order 3 & 0.2228371 & 0.2166688 & 326.2303 & 0.02854527 & 0.01036279 & 282.4505 \\ \hline
    Random Forest & 0.8747755 & 0.8724317 & 121.7878 & 0.7588743 & 0.7543612 & 140.719 \\ \hline
    \end{tabular}
  \end{table*}

\section{Results}

  \subsection{Regression Analysis Findings:}

    \subsubsection{Multivariate Linear Regression:} The linear regression model revealed that location variables, particularly proximity to city centers and major tourist attractions, significantly impact pricing. Room type and property size were also notable determinants.
    Basic MLR, MLR with Stepwise Selection (Step), and MLR with Interaction Variables (IVs) all show moderate performance, with Train R² values around 0.215 and Test R² values around 0.337. This indicates a modest fit to the data.
    The introduction of stepwise selection and interaction variables does not significantly improve the model's predictive accuracy, as indicated by very similar R² and Root Mean Square Error (RMSE) values across these variations.
    \subsubsection{Polynomial Regression:} This model highlighted non-linear relationships, showing that factors such as the number of amenities and host reputation have a varying impact on pricing at different levels.
    Polynomial models with Order 2 and Order 3, with and without stepwise selection and interaction variables, show similar performance to MLR in terms of R² and RMSE.
    The polynomial model with Order 3 and interaction variables demonstrates a noticeable improvement in the training phase (Train R² = 0.233), but a decrease in performance in the test phase (Test R² = 0.190), suggesting potential overfitting.
    \subsubsection{Lasso Regression:} The Lasso regression model with Order 1 shows similar performance to MLR, with Train R² = 0.215 and Test R² = 0.337. This indicates that the model's predictive accuracy is comparable to the linear regression model.
    \subsubsection{Interaction Effects:} The analysis of interaction variables indicated complex interdependencies, such as the combined effect of location and property type on pricing.

  \subsection{Random Forest Model Results:}
  In the analysis of Airbnb pricing dynamics within major European cities, the results obtained from the random forest regression model provided pivotal insights, particularly emphasizing the multifaceted nature of the factors influencing pricing.

  The model's outcome highlighted host-related variables, such as the response rate and the number of listings managed, as significant determinants of rental prices. This revelation underscores the impact of host engagement and professional experience in the pricing structure, a perspective that adds depth to the conventional understanding of rental pricing determinants.

  Geographical factors, especially the coordinates of latitude and longitude, emerged as crucial in the price determination process. This result corroborates the intuitive understanding that a property's specific location within a city significantly influences its value in the rental market. Furthermore, the attraction index was identified as the most influential variable, suggesting a direct correlation between a property's proximity to popular tourist destinations and its pricing. This finding aligns with the expectation that areas with higher tourist appeal tend to attract higher rental prices due to increased demand.

  When compared to other regression models employed in the study, the random forest model exhibited superior performance. It achieved an Adjusted 
  $R^2$ value of 0.7543 and an $RMSE$ of 140.719 on the test dataset, markedly outperforming the best linear regression model, which recorded an Adjusted 
  $R^2$ value of 0.3377 and an $RMSE$ of 233.1563. This significant improvement in predictive accuracy and explanatory power underscores the effectiveness of the random forest approach in capturing the complex interplay of variables influencing Airbnb pricing.

In conclusion, the findings from the random forest model have significantly advanced our understanding of the determinants of Airbnb pricing in Europe. They reveal a complex interaction of host characteristics, geographical location, and tourist attractions in shaping rental prices, providing a comprehensive perspective that extends beyond traditional pricing models.

  \begin{table}[ht]
    \centering
    \caption{Feature Importance Based on IncNodePurity}
    \label{tab:feature_importance}
    \begin{tabular}{|l|c|}
    \hline
    \textbf{Feature} & \textbf{IncNodePurity} \\ \hline
    room\_type                  & 122087850     \\ \hline
    host\_is\_superhost          & 26594131      \\ \hline
    multi                      & 21575815      \\ \hline
    biz                        & 24190212      \\ \hline
    city\_day                   & 114299029     \\ \hline
    person\_capacity            & 183757090     \\ \hline
    cleanliness\_rating         & 79985960      \\ \hline
    guest\_satisfaction\_overall & 162106050     \\ \hline
    bedrooms                   & 342017315     \\ \hline
    dist                       & 423128244     \\ \hline
    metro\_dist                 & 248206656     \\ \hline
    attr\_index\_norm            & 662153431     \\ \hline
    rest\_index\_norm            & 435838043     \\ \hline
    lng                        & 482308349     \\ \hline
    lat                        & 588248566     \\ \hline
    \end{tabular}
  \end{table}

  \begin{figure}[htbp]
    \centering
    \includegraphics[width=0.5\textwidth]{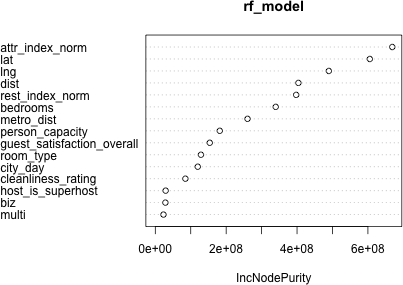}
    \caption{}
    \label{fig:my_label}
  \end{figure}

  \subsection{Model Comparison and Evaluation:}
  The comparative analysis of the models' performance metrics clearly indicates that the random forest model surpasses the linear and polynomial regression models in terms of explanatory power and predictive accuracy.

  \textbf{R-squared Comparison:} The R-squared value for the random forest model stands at an impressive 0.7588743 in the testing phase, as per Table 1. This value is significantly higher than those observed for both linear and polynomial regression models. R-squared, a measure of the proportion of variance in the dependent variable that is predictable from the independent variables, thus indicates that the random forest model explains a much larger portion of the variability in Airbnb pricing compared to the other models.

  \textbf{Mean Squared Error (MSE) Analysis:} The MSE for the random forest model in the testing phase is notably lower, recorded at 140.719. MSE, which measures the average of the squares of the errors—i.e., the average squared difference between the estimated values and the actual value, is a critical metric for assessing predictive accuracy. A lower MSE for the random forest model confirms its superior performance in accurately predicting Airbnb prices, with fewer errors, compared to the regression models.

  In summary, the random forest model's high R-squared value coupled with a significantly lower MSE underscores its effectiveness in capturing the complex dynamics of Airbnb pricing in Europe. It provides a more reliable and accurate tool for stakeholders in the Airbnb market to predict prices, thus facilitating more informed decision-making.

\section{Discussion}

The findings of this study offer insightful contributions to the understanding of Airbnb pricing dynamics in Europe, revealing the multifaceted nature of factors influencing rental prices. The discussion below explores the implications of these results and their alignment with existing literature.

  \subsection{Interpretation of Findings:}
  The significant role of location variables, such as proximity to city centers and tourist attractions, is emphasized by the random forest model's high R-squared value of 0.7588743 in the testing phase, indicating strong explanatory power. This highlights the premium placed on convenience and accessibility in short-term rental markets.
  Property features, including room type and amenities, are also critical determinants of pricing, as evidenced by the random forest model's low test RMSE of 140.719. This suggests that such features are significant predictors of rental prices.
  The random forest model also sheds light on the importance of host-related factors, such as response rate and number of listings, indicating that credibility and reliability of hosts are key considerations for consumers.

  \subsection{Comparison with Existing Literature:}
  The emphasis on geographical variables is validated by the high performance of the random forest model. This demonstrates nuanced variations within urban areas, as evidenced by the model's predictive accuracy.
  The identification of non-linear relationships and interaction effects in the polynomial regression models extends the understanding of the complexity in Airbnb pricing strategies. The varying R-squared and RMSE values across different polynomial models highlight these intricate relationships.

  \subsection{Implications for Stakeholders:}
  Understanding these determinants, especially those highlighted by the predictive power of the random forest model, can aid hosts in optimizing pricing strategies to enhance profitability while remaining competitive.
  Travelers can use these insights to better understand pricing structures and make informed accommodation choices, considering the reliability of the model's predictions.
  Policymakers and urban planners can leverage this information to understand the economic impact of short-term rentals and guide regulatory measures.

  \subsection{Limitations and Future Research:}
  The study has limitations, including the scope of cities covered and potential biases in the dataset. The varying performance of the models suggests room for expanding the dataset.
  Future research could explore more diverse geographical areas and incorporate longitudinal data to capture temporal pricing trends. Investigating the impact of external factors, such as local regulations and economic conditions, on Airbnb pricing would also be valuable.

  \subsection{Concluding Remarks:}
  The use of advanced machine learning techniques, particularly the random forest model with its high R-squared and low RMSE values, provides a deeper understanding of the shared economy’s pricing mechanisms. This study offers valuable insights into the multifaceted nature of rental pricing in the Airbnb market.

\section{Conclusion}

This study embarked on an analytical journey to unravel the complex dynamics of Airbnb pricing in major European cities. Through the application of advanced machine learning techniques, including regression analysis and random forest models, it has illuminated the multifaceted nature of factors influencing short-term rental prices.

Key conclusions drawn from the research are as follows:

  \subsection{Significant Determinants of Airbnb Pricing:} The study confirmed the critical role of location-specific factors, such as proximity to city centers and tourist attractions, in determining rental prices. Additionally, property characteristics like room type and amenities, along with host-related attributes, were identified as significant influencers.

  \subsection{Methodological Contributions:} The use of sophisticated data analytics tools provided a comprehensive understanding of the pricing mechanisms. The superiority of the random forest model in capturing the complexity of the dataset highlights the utility of machine learning in real-world economic analysis.

  \subsection{Practical Implications:} For hosts on Airbnb, the insights offer guidance on optimizing rental prices. For travelers, the study provides a lens to understand and predict rental pricing. Policymakers and urban planners can also draw valuable information for regulating and understanding the impact of short-term rentals.

  \subsection{Limitations and Future Research:} Acknowledging the study's limitations, such as the geographical scope and potential dataset biases, future research directions include expanding the analysis to more diverse locations and incorporating temporal data to observe pricing trends over time. Additionally, exploring the impact of external factors like local regulations and economic conditions could yield further valuable insights.

  \subsection{Concluding Thoughts:} Ultimately, this research contributes significantly to the understanding of Airbnb pricing strategies, offering a nuanced view of the variables that shape them. It stands as a testament to the power of data science in elucidating complex market phenomena and lays the groundwork for future explorations in the shared economy domain.






%

\newpage
\setcounter{section}{0} 
\section*{Appendix}
\section{Data}

Each major city has its own dataset for weekend and weekdays. Variables included in the dataset are:

\begin{itemize}
  \item Host\_ID (Id)
  \item Total price of listing (realSum)
  \item Room type: private, shared, entire home, apartment (room\_type)
  \item Whether or not the room is shared (room\_shared)
  \item Max number of people allowed on the property (person\_capacity)
  \item Host's Superhost status (host\_is\_superhost): Indicates whether the host is recognized as a superhost.
  \item Whether it is multiple rooms (multi)
  \item Whether for business or family use (biz)
  \item Distance from the city center (dist)
  \item Distance from the nearest metro (metro\_dist)
  \item Latitude and longitude (lat, lng)
  \item Guest satisfaction (guest\_satisfaction\_overall)
  \item Cleanliness (cleanliness\_rating)
  \item Total quantity of bedrooms available among all properties for a single host (bedrooms)
  \item Index of attractions near the hotel (attr\_index)
  \item Normalized Index of Attractions near the hotel (attr\_index\_norm)
  \item Index of Restaurants near the hotel (rest\_index)
  \item Normalized Index of Restaurants near the hotel (rest\_index\_norm)
\end{itemize}

The dataset consists of:

\begin{itemize}
  \item Continuous variables:
    \begin{itemize}
      \item realSum, dist, metro\_dist
      \item lat, lng
      \item attr\_index, attr\_index\_norm
      \item rest\_index, rest\_index\_norm
    \end{itemize}
  \item Ordinal variables:
    \begin{itemize}
      \item person\_capacity
      \item guest\_satisfaction\_overall
      \item cleanliness\_rating
      \item bedrooms
    \end{itemize}
  \item Nominal variables:
    \begin{itemize}
      \item room\_type, room\_shared
      \item host\_is\_superhost
      \item multi, biz
    \end{itemize}
\end{itemize}

\end{document}